\documentclass[runningheads]{llncs}
\usepackage{graphicx}
\usepackage{multirow}
\usepackage{subcaption}
\usepackage{amsmath,amssymb}
\usepackage[ruled,vlined]{algorithm2e}

\usepackage[dvipsnames]{xcolor}

\begin{document}
\title{Learning Generative Deception Strategies in Combinatorial Masking Games}
\author{Junlin Wu\inst{1} \and
Charles Kamhoua\inst{2} \and
Murat Kantarcioglu\inst{3} \and
Yevgeniy Vorobeychik\inst{1}
}
\authorrunning{Junlin Wu et al.}
\institute{Washington University, Saint Louis, MO 63130, USA 
\email{\{junlin.wu,yvorobeychik\}@wustl.edu}
\and
Army Research Laboratory, Adelphi, MD 20783, USA
\email{charles.a.kamhoua.civ@mail.mil}\\
\and
University of Texas, Dallas, TX 75080, USA\\
\email{muratk@utdallas.edu}
}
\maketitle              %
\begin{abstract}
Deception is a crucial tool in the cyberdefence repertoire, enabling defenders to leverage their informational advantage to reduce the likelihood of successful attacks.
One way deception can be employed is through obscuring, or masking, some of the information about how systems are configured, increasing attacker's uncertainty about their targets.
We present a novel game-theoretic model of the resulting defender-attacker interaction, where the defender chooses a subset of attributes to mask, while the attacker responds by choosing an exploit to execute.
The strategies of both players have combinatorial structure with complex informational dependencies, and therefore even representing these strategies is not trivial.
First, we show that the problem of computing an equilibrium of the resulting zero-sum defender-attacker game can be represented as a 
linear program with a combinatorial number of system configuration variables and constraints, and develop a constraint generation approach for solving this problem.
Next, we present a novel highly scalable approach for approximately solving such games by representing the strategies of both players as neural networks.
The key idea is to represent the defender's mixed strategy using a deep neural network generator, and then using alternating gradient-descent-ascent algorithm, analogous to the training of Generative Adversarial Networks.
Our experiments, as well as a case study, demonstrate the efficacy of the proposed approach.

\end{abstract}
\section{Introduction}

The use of deception in cyber defense has a long tradition. Honeynets are perhaps the most popular examples~\cite{Dagon04,Provos03,Spitzner03}, but numerous other ideas, such as adding fake information or changing observable configurations of machines or networks have also been explored~\cite{Albanese16,Rowe07,Schlenker18,Shi20,Wang18}.
While many such studies have either focused on lower-level implementation issues or qualitative analysis, there has emerged a robust literature that formally models deception as a game-theoretic interaction between a defender and an attacker~\cite{Guo17,Horak17,Kiekintveld15,Nguyen19,Pibil12,Schlenker18,Shi20,Wang18,Yin14}.

While game-theoretic models of deception are fundamentally appealing, as they aspire to use deploy such tools even while accounting for highly sophisticated adversaries who carefully reason about it, approaches are typically either highly stylized and qualitative~\cite{Carrol11,Horak17,Pawlick15}, or use game representations that do not scale well with dimensionality of system configuration space~\cite{Guo17,Schlenker18,Wang18}.
For example, Schlenker et al.~\cite{Schlenker18} represent the set of possible system attributes and possible deceptions (which, in this work, involve observable characteristics of the systems) by enumerating all possibilities.
This representation, however, is exponential in the number of system attributes, and real systems may have hundreds of these.
Furthermore, much of recent work on deception presumes that the attacker's decision amounts to the choice of a target system to attack.
In real cybersecurity encounters, attacks are launched via exploits, and once an exploit is developed, it can in principle be used nearly indiscriminately against \emph{any machine which has the vulnerable operating system, applications, required open ports, and so on}.
To the extent that exploit deployment against a particular organizational network is automated, \emph{all} vulnerable machines can be targeted simultaneously.

We build on the insight offered by Shi et al.~\cite{Shi20}, who represent systems that are to be defended using a collection (vector) of features.
Deception in their model entails modifications to individual features, subject to linear constraints.
Unlike this, and much of other prior work, however, we turn our attention to a relatively underinvestigated means for deception through \emph{masking}, rather than changing true system attributes.
As shown recently, masking can be just as effective as changing features~\cite{Estornell20}, but is in practice often easier to implement.

Specifically, we introduce a \emph{combinatorial masking game (CMG)}, in which a defender controls a collection of potentially vulnerable computers, each characterized by a vector of features (e.g., OS type and version, applications installed and their versions, etc.).
The defender (whom we call Alice) chooses a subset of features to mask for each machine.
The attacker (whom we name Bob), in turn, observes the non-masked features of all machines, as well as which features are masked and which are not, and in response chooses an exploit to execute against the \emph{entire collection of defender's computers}.
Consequently, all of the computers which contain the exploited vulnerability are viewed as successfully compromised, and the attacker gains (while the defender loses) the total value of these (which is, in general, a function of their \emph{true} features).

The game above is a zero-sum Bayesian game, and we seek its Bayes-Nash equilibrium which characterizes both the mixed strategies of the defender (i.e., the randomized deception strategy) and of the attacker.
Our first step is to derive a linear programming (LP) formulation of the Bayes-Nash equilibrium solution of this game.
Unfortunately, the resulting LP is even intractable to represent as we increase the number of features, since the strategies of both players are combinatorial in size.
Our solution is to represent the strategies of both players as neural networks.
This is straightforward for the attacker, as he chooses among a set of exploits, which we explicitly enumerate.
For the defender, however, we need to represent a probability distribution over all possible masking strategies---a set combinatorial in the feature space dimension.
Adding a constraint generation procedure helps, but the approach still fails to scale beyond tiny problem instances.
Our solution is to represent these as \emph{generative neural networks}, akin to the generator in Generative Adversarial Networks (GANs)~\cite{Goodfellow14}.
Since both strategies are now differentiable, we develop a gradient descent-ascent algorithm for learning these (resulting in an approximate Bayes-Nash equilibrium), which is inspired by the algorithm used for training GANs (although the specifics of the training process, such as the loss function, are quite different).

Finally, we evaluate the proposed approach experimentally.
First, we show that our approach is near-optimal (compared to linear programming) on small problem instances, with significantly better scalability.
Next, we compare it with three baselines: random masking, unconditional masking, (independent of actual device configuration), and a heuristic greedy approach for masking.
We show that while random masking is extremely fast, it typically results in solutions that are much worse for the defender.
Greedy heuristic yields better solutions, but our approach still offers a significant improvement over this approach, and is in fact also much more scalable.
Proposed approach also yields better solutions than unconditional masking.
We close with a case study of a  synthetic example that illustrates the nature of our solutions.

\section{Related Work}

One of the most common concrete instantiations of deception are honeypots and honeynets~\cite{Dagon04,Provos03,Spitzner03}.
One of the main ideas behind honeypots is to detect and investigate cyber threats, taking advantage of the information asymmetry that favors the defender, who knows which of their machines are real and which are honeypots, in contrast to the attackers who, at least in theory, do not.
This idea has a number of variations, such as adding ``honey'' (fake) accounts and fake data~\cite{Bercovitch11,Juels13,Abay2019}.

One of the early abstractions of cyber deception was proposed by Cohen, who studied it as a problem of guiding attackers through a benign part of an attack graph~\cite{Cohen01a,Cohen01b,Cohen03}.
A further formalization of this idea was to investigate how deception can impact the evolution of the attacker's beliefs~\cite{Estornell20,Horak17}.

One of the earliest game-theoretic modeling approaches to deception was through \emph{signaling games}, in which a defender (sender) has a type that they may deceptively communicate (signal) to an attacker (receiver)~\cite{Carrol11,Pawlick15}.
However, these models were relatively abstract and simplistic.
An alternative paradigm of security games, in which the interactions between a defender, who protects a set of targets, and an attacker, who chooses the best target to attack, provided a higher-resolution game-theoretic modeling framework for studying strategic security interactions~\cite{Tambe11}.
This framework then gave birth to some of the most recent investigations of deceptive signaling in security, leveraging the defender's informational advantage about which targets have been chosen to be protected (which is only observed by the attacker after they choose the target to attack)~\cite{Rabinovich15,Yan20,Xu15}.

Several recent game-theoretic models for deception provide the core intellectual precedent for our work.
Schlenker et al.~\cite{Schlenker18} introduced the idea of observable configurations as the defender's strategy space, with the attacker choosing a target to attack after reasoning about the posterior distribution of actual, given observable, configurations.
The key technical limitation in that work is the requirement of fully enumerating the entire configuration space (both actual and observed) in the model.
Shi et al.~\cite{Shi20} address this limitation by proposing a factored (feature-based) representation of these, but the game-theoretic model they use involves a myopic bounded-rational attacker who does not explicitly reason about deception.
We build on both of these, using both a feature-based representation of the problem that enables us to take algorithmic advantage of problem structure, but at the same time model attackers as fully rational---that is, fully reasoning about deception.

Our approach of using generative neural networks to represent defender's mixed strategies is partly inspired by Generative Adversarial Networks (GANs)\\~\cite{Goodfellow14} and conditional GANs~\cite{Mirza14}, as well as the use of such representations in fictious play algorithms for solving games~\cite{Kamra19}.
Our key idea is to use conditional GANs that are a function of a true configuration to learn an implicit mixed strategy for the defender.
This is also quite unlike Kamra et al., who learn unconditional generative model as a randomized \emph{best response} to a fixed memory of actions by the other players.

\section{Deception through Attribute Masking}
\label{S:model}

Consider a defender (Alice) in charge of security for an organizational network comprised of a collection of $m$ devices.
Each device is characterized by a feature vector of attributes
$x = (x_i)_{i\in[n]}$, where $[n]$ denotes the set $\{1,\ldots,n\}$.
Each attribute, $x_i$, in turn, can take on one of a finite collection of values, i.e., ${x_i \in X_i \subseteq \{-1,1,\ldots,V\}}$, where the attribute value of $-1$ corresponds to a default configuration (e.g., application is not installed, port is not open) or to ``N/A'' (e.g., a version number of an application that is not installed), and $V$ the largest possible attribute value.
When there are multiple devices on the network, we represent each device by $x^k$.
However, we will omit this superscript when it is either not relevant, or not important.

This defender faces an attacker (Bob) who aspires to compromise as many of these devices as he can.
More precisely, let $v(x)$ be the value to the attacker successfully compromising a device with  configuration $x$; we assume that $v(x)$ is also the loss to the defender in the event of compromise.
If $S$ is a set of devices the attacker successfully compromises, the resulting utility of the attacker (and loss to the defender) is then $\sum_{k \in S} v(x^k)$.

The means that the attacker uses for his ends is to choose an exploit $e$ from a collection of actionable exploits $E$.
The set $E$ can be alternatively viewed as a collection of exploitable vulnerabilities, and the attacker chooses one of these to develop a custom exploit for, leveraging any additional information about the target network.
We assume that the attacker chooses only a single exploit from this collection.
Each exploit $e \in E$ is associated with a set of configurations that the exploit requires to successfully execute.
We assume that this set, which we denote by $X^e$, can be specified as a conjunction of required sets for each attribute, that is,
$X^e = \{X^e_1,\ldots,X^e_n\}$, where $X^e_i = [a_{i1},\ldots,a_{il_i}] \subseteq X_i$.
The interpretation is that the value of each attribute $x_i$ must be in the set $X^e_i$ in order for the exploit $e$ to successfully execute.
For example, an exploit may target all versions of a Chrome browser between versions 75 and 85 installed on Windows  10 versions 1500-1900, as long as ports 23 and 25 are open.
Note that the values of most attributes may not be relevant to attack execution, in which case $X^e_i = X_i$.
We use notation $x \in X^e$ to mean that configuration $x$ satisfies the requirements of the exploit $e$ and, consequently, the device with this configuration can be compromised by $e$.

Once an exploit $e$ is chosen by Bob, \emph{all} the devices on Alice's network which can be successfully attacked by it are compromised.
The gain to Bob, and loss to Alice, is then $\sum_{k=1}^m v(x^k)\delta(x^k \in X^e)$, where $\delta(\cdot)$ is an indicator function which is 1 if the condition is True and 0 otherwise.

To deal with this predicament, Alice (the defender) can mask a subset of configuration attributes of her devices.
Let $y^k \in \{0,1\}^n$ denote this mask applied to device $k$, where $y_i^k = 0$ means attribute $i$ of device $k$ is suppressed (not observable) and $y_i^k = 1$ means that it can be observed (by Bob, the attacker, as well as, potentially others).  
Thus, we only allow suppression of attributes, but not changing their observed values as done in prior work~\cite{Schlenker18,Shi20}.
In addition, and crucially, we assume that the masked attributes cannot be easily inferred from the observed ones (except by the attacker computing a posterior, as discussed below). 
Of course, masking is costly for a number of reasons. 
For example, information about attributes can be important to broadcast to ensure proper implementation choices and application compatibility.
We let $c(y^k)$ denote the cost of choosing a mask $y^k$ for a device $k$.
We assume that the total masking cost is additive over devices, that is, $c(y^1,\ldots,y^m) = \sum_k c(y^k)$.

Given a true configuration $x^k$ and a mask $y^k$ for a device $k$, the attacker observes two things: 1) the mask $y^k$ (inability to see the particular attributes of the device gives it away) and 2) the true values of the \emph{observable} attributes, which we denote by $\tilde{x}^k = x^k \odot y^k$, where $\odot$ is a Hadamard product.
Indeed, note that $\tilde{x}^k$ actually captures all of the relevant information, since in our notation above, $\tilde{x}_i^k = 0$ necessarily implies that attribute $i$ is not observed (since observed values do not include 0 in our problem encoding).
This notation will prove convenient below.

The game which we described above, which we call a \emph{Combinatorial Masking Game (CMG)}, constitutes a Bayesian game in which $(x^k)$ (the actual configurations of the devices) is private information of the defender, while the attacker observes $(\tilde{x}^k)$, observable features after masks $y^k$ have been applied to all devices $k$.
Let $p(x^1,\ldots,x^m)$ be the prior distribution over device configurations on the network, which is common knowledge to both Alice and Bob.
As noted above, the utility (after all uncertainty is resolved) of both players depends \emph{only} on the configurations $x^k$ of the defender's devices and the exploit chosen by the attacker $e$, but \emph{not Alice's masking choices} $y^k$, which serve solely as a means of deception.
Since this is a Bayesian game, the defender's mixed strategy is a probability distribution over masks $y^k$ conditional on actual configurations $x^k$.
Letting $\mathbf{y}$ and $\mathbf{x}$ be the vectors that concatenate the masks chosen by the defender for all devices and the actual device features, respectively, we formally denote her mixed strategy by $q(\mathbf{y};\mathbf{x}) = \Pr\{\mathbf{y}|\mathbf{x}\}$.
The attacker's mixed strategy, in turn, is the probability of choosing an exploit $e$ given his observation of the devices $\tilde{x}^k$, which we concatenate into a vector $\mathbf{\tilde{x}}$.
Formally, we denote this by $z(e;\mathbf{\tilde{x}}) = \Pr\{e|\mathbf{\tilde{x}}\}$.

We denote by $u(q,z)$ the expected utility of the attacker choosing a mixed strategy $z$ while the defender chooses $q$. 
Our goal is to compute a (mixed-strategy) Bayes-Nash equilibrium (BNE) of this zero-sum game.
In our setting, a strategy profile $(q^*,z^*)$ is a BNE if
\[
q^* \in \arg\min_q \left(u(q,z^*) + \mathbb{E}_{\mathbf{y} \sim q}[c(\mathbf{y})]\right) \ \mathrm{and} \
z^* \in \arg\max_z u(q^*,z).
\]
Note that since this game is strategically zero-sum, the BNE strategy $q^*$ of Alice is also her Stackelberg equilibrium strategy~\cite{korzhyk2011stackelberg}.

\section{Computing Equilibrium Deception Strategies}
\label{S:computation}

Recall that our goal is to compute a BNE of the game presented in Section~\ref{S:model}.
We begin our discussion of BNE computation in \emph{CMGs} by considering a single device in the charge of the defender.
In Section~\ref{S:multipledevices} we extend the approach to an arbitrary collection of such devices.
Since we are dealing with a single device, we omit the superscripts $k$ throughout this section.

Central to our task will be to derive the precise expressions for the best responses of both the attacker and defender.
These expressions will subsequently naturally lead to a linear programming representation of our problem, which in turn yields the first (but highly intractable) solution approach.
We begin by deriving an expression for the attacker's best response problem.

\subsection{Computing the Attacker's Best Response}

Consider a defender who plays a mixed strategy $q(y;x)$, where $x$ is the true feature vector for the (single) device, while $y$ is the associated mask, and $\tilde{x}$ is the  feature vector for the device observed by the attacker.
We now derive an expression for the attacker's best response to this strategy.

The first step is to obtain the attacker's posterior distribution over the device configuration $x$ given observation $\tilde{x}$ (where we explicitly use both $\tilde{x}$ and $y$ as observations for clarity):
\begin{equation*}
    b(x;\tilde{x}, y) \equiv \Pr\{x|\tilde{x}, y\} = \frac{\Pr \{\tilde{x}, y|x\}p(x)}{p(\tilde{x}, y)},
\end{equation*}
where $p(\tilde{x}, y) = \sum_x \Pr\{\tilde{x}, y|x\}p(x)$.
Now, note that
\begin{equation*}
\Pr\{\tilde{x},y|x\} = \Pr\{\tilde{x}|y,x\}\Pr\{y|x\} = \Pr\{\tilde{x}|y,x\} q(y;x) = \delta(\tilde{x} = x \odot y) q(y;x).
\end{equation*}
Based on the definition of $\tilde{x}$, if $\tilde{x} = x \odot y$, then $\Pr\{\tilde{x}|y,x\} = 1$, and otherwise, it is 0. Thus, we can represent $\Pr\{\tilde{x}|y,x\}$ using the indicator function $\delta(\mathit{Cond})$ where $\delta(\mathit{Cond}) = 1$ if $\mathit{Cond}$ is true, and 0 otherwise.

Since a successful attack on the device with configuration $x$ yields the attacker a value $v(x)$ which is lost to the defender, the utility of the attacker for deploying exploit $e$ after observing $(\tilde{x}, y)$ is
\begin{align*}
u_a(e,\tilde{x},y,q) &= \sum_x v(x) b(x;\tilde{x}, y) \delta(x \in X^e) \\
&= \frac{1}{p(\tilde{x}, y)}\sum_x v(x)q(y;x)p(x)\delta(x \in X^e)\delta(\tilde{x} = x \odot y).
\end{align*}

Next, recall that $z(e;\tilde{x})$ represents the attacker's mixed strategy, that is, the probability distribution over exploits $e$ chosen.
Moreover, it is important to keep in mind that $\tilde{x}$ is (implicitly) a function of $y$, which is observed by the attacker, as well as $x$, which is not.
The attacker's optimal utility is then
\begin{equation}
    \label{E:brattaacker2}
    u_a^*(\tilde{x},y,q) = \max_z \sum_e z(e;\tilde{x}) u_a(e,\tilde{x},y,q),
\end{equation}
that is, this is the maximum utility that the attacker achieves by choosing an optimal exploit to deploy against the defender's device.

Finally, we will use a mathematical trick to rewrite the attacker's best response condition in a form that will prove more convenient. Note that mathematically, it makes no difference if we optimize $z$ separately for each $(\tilde{x},y)$, or simultaneously over all $(\tilde{x},y)$ where we maximize \emph{expected utility} with respect to the prior distribution $p(\tilde{x},y)$ over configurations. Thus, for the attacker, the maximization problem in Equation (\ref{E:brattaacker2}) is equivalent to
\begin{subequations}
\begin{align}
    u_a^*(q) &= \max_z \sum_{\tilde{x},y} p(\tilde{x},y) u_a^*(\tilde{x},y,q) \\
    &= \max_z \sum_{\tilde{x},y} \sum_e z(e;\tilde{x}) \sum_x v(x)q(y;x)p(x)\delta(x \in X^e)\delta(\tilde{x} = x \odot y)\\
    &= \max_z \sum_x p(x)\sum_{y} q(y;x) \sum_e z(e;\tilde{x})  v(x)\delta(x \in X^e),
        \label{E:brattaacker3}
        \end{align}
        \end{subequations}
where $\delta(\tilde{x} = x \odot y)$ and the sum over $\tilde{x}$ are no longer necessary, since we are already summing over $x$ and $y$ and the terms where $\tilde{x} \neq x \odot y$ will yield 0.

\subsection{Computing the Defender's Best Response}

We now turn to deriving a similar expression for the defender's best response to an attacker's mixed strategy $z(e;\tilde{x})$.

For the defender, who knows $x$, chooses $y$, and faces an attack $e$, the utility is
\begin{equation*}
    u_d(e,y;x) = -(v(x)\delta(x \in X^e)+c(y)).
\end{equation*}

Since the defender actually chooses a randomized strategy $q(y;x)$ and aims to maximize the utility over all such strategies $q$ in response to the attacker's mixed strategy $z(e;\tilde{x})$, the optimal expected utility for the defender is
\begin{equation}
    u_d^*(z;x) = - \min_q \sum_y q(y;x) \left( \sum_e z(e;\tilde{x})v(x)\delta(x \in X^e)+c(y)\right).
    \label{E:brdefender}
\end{equation}
Moreover, maximizing the defender's utility for a given $x$ is equivalent to maximizing the expected utility with respect to the prior distribution $p(x)$. Thus, we can redefine the defender's ex ante utility as follows:
\begin{subequations}
\begin{align}
u_d^*(z) &= -\min_q \sum_x p(x) u_d^*(z;x)\\
&= -\min_q \sum_x p(x)\sum_y q(y;x) \left(\sum_e z(e;\tilde{x})v(x)\delta(x \in X^e)+c(y)\right)
\label{E:brdefender2}
\end{align}
\end{subequations}

\subsection{Computing Equilibrium Deception}

Recall that the pair of strategies $(q,z)$ constitute a (Bayes-)Nash equilibrium iff they jointly satisfy Equations \eqref{E:brattaacker3} and \eqref{E:brdefender2}.
Since this game is zero-sum, BNE deception strategy and Bayes-Stackelberg equilibrium deception coincide, and we consequently focus on computing a BNE deception strategy for the defender (the attacker's equilibrium strategy ultimately serves as a means to that end).

We can rewrite the BNE of the deception game as the following minimax problem:
\begin{equation}
   \min_{q}\max_{z} \quad \sum_x p(x) \sum_y q(y;x)\left(\sum_e z(e;\tilde{x})v(x)\delta(x \in X^e) + c(y)\right). 
   \label{E:NashEq}
\end{equation}
This, in turn, can be represented as the following linear program (LP):
\begin{subequations}
\label{E:LP}
\begin{align}
&\min_{q \ge 0,u_a^*} \quad u^*_a + \sum_x p(x) \sum_y q(y;x) c(y)\\
&\mathrm{s.t.:}\\
&u_a^* \ge \sum_x p(x) \sum_y q(y;x)\left(\sum_e z(e;\tilde{x})v(x)\delta(x \in X^e)\right) \quad \forall \ z(e;\tilde{x})\label{C:attack}\\
&\sum_y q(y;x) = 1 \quad \forall \ x\\
&\sum_e z(e;\tilde{x}) = 1 \quad \forall \ \tilde{x}, y.
\end{align}
\end{subequations}
Note that here, the Constraints~\eqref{C:attack} are for all possible attack strategies (i.e., functions of $\tilde{x}$).
However, since there is always a pure strategy best response, we can restrict this to consider only \emph{deterministic} attack strategies.
Nevertheless, the set of constraints is exponential in possible $\tilde{x}$, in addition to the fact that the number of variables in this LP is exponential (ranging over the entire domains of $x$ and $y$).
Consequently, even though we can use standard tools, such as CPLEX, to solve this LP in principle, scalability will be severely limited.

\begin{algorithm}[h!]
\SetAlgoLined
\KwIn{Exploits set $E = \{e_1, e_2, \cdots\}$; $p(x)$; cost function $c$}
\KwOut{Optimal utility for defender and attacker; defender's optimal strategy $q(y;x)$; attacker's optimal strategy $z(e;\tilde{x})$.}
Initialization: randomly generate some attacker's strategy $\{z(e;\tilde{x})\}$ set $Z$; $err \xleftarrow{} \infty$; tolerance $\epsilon$\;
\While{err $>$ $\epsilon$}{
1. Solve defender's LP:
\begin{subequations}
\begin{align*}
&\min_{q \ge 0,u_a^*} \quad u^*_a + \sum_x p(x) \sum_y q(y;x) c(y)\\
&\mathrm{s.t.:}\\
&u_a^* \ge \sum_x p(x) \sum_y q(y;x)\left(\sum_e z(e;\tilde{x})v(x)\delta(x \in X^e)\right) \quad \forall \ z(e;\tilde{x}) \in Z\\
&\sum_y q(y;x) = 1 \quad \forall \ x.
\end{align*}
\end{subequations}

2. Fix $q(y;x)$ from defender's LP solution and solve the attacker's LP:
\begin{subequations}
\begin{align*}
&\max_{z \in \{0,1\}} \quad \sum_x p(x) \sum_y q(y;x)\left(\sum_e z(e;\tilde{x})v(x)\delta(x \in X^e)\right)\\
&\mathrm{s.t.:}\\
&\sum_e z(e;\tilde{x}) = 1 \quad \forall \ \tilde{x}, y.
\end{align*}
\end{subequations}

3. Add attacker's LP solution $\{z(e;\tilde{x})\}$ to $Z$

4. Calculate $err$ $\xleftarrow{}$ Abs(defender's LP obj - $\sum_x p(x) \sum_y q(y;x) c(y)$ - attacker's LP obj)
}
 \caption{Constraint generation algorithm for solving the linear programming.}
 \label{lpalgo}
\end{algorithm}

To partially address the scalability challenge, we can use constraint generation to avoid explicitly enumerating Constraints~\eqref{C:attack} corresponding to possible attacks.
Algorithm \ref{lpalgo} formalizes this approach, which at the high level proceeds as follows.
We start with a small set of constraints (attacker strategies), solve the resulting relaxed LP, then compute the attacker's best response, which is added to the LP, and the process is then repeated until convergence.
Note that although we still need to enumerate the attacker strategies in computing the best response, we avoid the key bottleneck, which is \emph{space} complexity (having to explicitly represent the LP with all of the constraints in memory is a greater bottleneck than enumeration of these).

Although using constraint generation can significantly reduce the size of the LPs we have to store in memory, it will still scale poorly in the dimensionality $n$ of the feature representation space of the devices.
Next, we describe our approach for entirely side-stepping the scalability challenge by representing the defender and attacker mixed strategies as neural networks, and then solving the game using a gradient-based method.

\subsection{Scalable Approximation of Equilibrium Deception through Generative Adversarial Masking}

To solve \emph{CMGs} at scale, we now propose a novel gradient-based learning method inspired by generative adversarial networks (GANs), which we term \emph{generative adversarial masking (GAM)}.
The key idea is to first represent the strategies of both players using deep neural networks, and then leverage an alternating gradient descent-ascent algorithm with the defender's expected loss as the objective.

To begin, we rewrite Equation~\eqref{E:NashEq} in a manner that will prove especially convenient.
Specifically, note that this expression is equivalent to first taking the expectation with respect to $x \sim p(x)$ (i.e., $x$ distributed according to the prior distribution $p(x)$), and then taking the expectation with respect to $y \sim q(y;x)$, where the distribution is actually defined by the defender's mixed strategy, and conditional on $x$.
We thus rewrite Equation~\eqref{E:NashEq} as follows:
\begin{equation}
    \min_{q} \max_{z} \quad \mathbb{E}_{x\sim p(x)}
\mathbb{E}_{y \sim q(y;x)} \left( \sum_e z(e;\tilde{x}) v(x)\delta(x \in X^e) + c(y)\right).
\label{E:min-max}
\end{equation}
Now, suppose we represent the attacker's strategy $z(e;\tilde{x})$ as a deep neural network with parameters $\theta$, i.e., $z(e;\tilde{x};\theta)$.
Of course, we need to ensure that this is a valid probability distribution over $E$, but that is straightforward to implement by adding a softmax layer, just as in standard classification problems.
The strategy of the attacker is then simply a parametric function, with parameters $\theta$, that takes $\tilde{x}$ as input and outputs a distribution over $e$, as desired.

The representational idea above does not, however, work for the defender, as it is inherently intractable to \emph{explicitly} represent an arbitrary probability distribution over $y$ (since the number of outputs becomes exponential).
Instead, we propose to use a \emph{conditional generative neural network (CGNN)} (or simply \emph{generator}) as an \emph{implicit} representation of this distribution, as is done in GANs.
A CGNN takes two inputs: 1) the conditioning input $x$ (which in our case is the true device configuration), and 2) a random variable $r \in [0,1]^n$, which we assume is distributed uniformly at random.
We write the resulting CGNN representation  as (deep neural network) $Q(x,r;\beta)$, where $\beta$ are the neural network parameters.
For a given input $(x,r)$, the CGNN \emph{deterministically} outputs $y$; consequently, since $r$ is generated stochastically, $Q$ induces a probability distribution over $y$ conditional on $x$.
Moreover, since $r$ is a valid probability distribution, so is $Q$.
By optimizing its parameters $\beta$, we can now optimize the probability distribution $Q$.
Since $y \in \{0,1\}^n$, we use sigmoid layer as the last layer of $Q(x,r;\beta)$
neural network and binarize $y$ every $k$ iterations ($k$ is a hyperparameter determined through experiment trials), as well as the last iteration using $0.5$ as the threshold to ensure the neural network can be properly trained and the final output $y$ generated by $Q$ is a binary vector.

Rewriting everything using both the CGNN $Q$ and the neural network for representing the attacker's best response $z$, we obtain
\begin{equation*}
\min_{\beta \ge 0} \max_{\theta \ge 0} \quad \mathbb{E}_{x\sim p(x)}
\mathbb{E}_{y \sim Q(x,r;\beta)} \left( \sum_e z(e;\tilde{x};\theta) v(x)\delta(x \in X^e) + c(y)\right).
\end{equation*}
The final useful observation is that the sole source of stochasticity in $y$ is the randomness of generated uniform random $r$, with $y = Q(x,r;\beta)$.
Consequently, we can rewrite as follows:
\begin{equation*}
\min_{\beta \ge 0} \max_{\theta \ge 0} \quad \mathbb{E}_{x\sim p(x)}
\mathbb{E}_{r \sim U^n} \left( \sum_e z(e;x \odot Q(x,r;\beta);\theta) v(x)\delta(x \in X^e) + c(Q(x,r;\beta))\right),
\end{equation*}
where $U^n$ is a uniform distribution over $[0,1]^n$.
Observe that now both expectations are unconditional, and so the order no longer matters.

To obtain the final algorithm for learning $\beta$ and $\theta$, we simply approximate the expectations using finite samples of $x$ and $r$, and alternate gradient descent (for updating $\beta$ and gradient ascent (for updating $\theta$) until convergence.
Algorithm~\ref{A:gda} presents the complete learning procedure.

\begin{algorithm}[h!]
\SetAlgoLined
\KwIn{Exploits set $E = \{e_1, e_2, \cdots\}$; $p(x)$; cost function $c$; number of samples $k$}
\KwOut{Optimal utility for attacker; optimal utility for defender; $(y^{(i)}, x^{(i)})$ which embeds the defender's optimal strategy $q(y;x)$; the attacker's optimal strategy $z(e;\tilde{x})$.}
 Initialization\;
 Sample $\{(x^{(1)},r^{(1)}), \cdots, (x^{(k)},r^{(k)})\}$ where $r \sim U^n$ and $x \sim p(x)$\;
 \For{number of training iterations}{
  \For{number of $N_z$ training}{
  Update $N_z$ by using gradient descent-ascent to maximize the objective $\nabla_{\theta_z}\frac{1}{k}\sum_{i=1}^k (\sum_e N_z(e;x^{(i)} \odot N_Q(x^{(i)},r^{(i)};\beta_Q);\theta_z)v(x^{(i)})\delta(x^{(i)}\in X^e))$\;
  }
  Update $N_Q$ by using gradient descent-ascent to minimize the objective $\nabla_{\beta_Q}\frac{1}{k}\sum_{i=1}^k (\sum_e N_z(e;x^{(i)} \odot N_Q(x^{(i)},r^{(i)};\beta_Q);\theta_z)v(x^{(i)})\delta({x^{(i)} \in X^e}) + c(N_Q(x^{(i)},r^{(i)};\beta_Q)))$\;
  }
 \caption{The \emph{GAM} gradient descent-ascent algorithm for training $N_Q$ and $N_z$ neural networks. $N_Q$ represents the neural network for $q(y;x)$ and $N_z$ represents the neural network for $z(e;\tilde{x})$.}
 \label{A:gda}
\end{algorithm}
\section{Extension to Multiple Devices}
\label{S:multipledevices}

In this section we extend the single-device approach presented in Section~\ref{S:computation} to a setting where the defender controls the security for multiple devices. 
Suppose that the defender has $m$ devices on the network whose configurations follow the distribution $\textbf{x} = (x^1, \cdots,x^m) \sim p(\textbf{x})$. 
The defender chooses the masking strategy $\textbf{y} = (y^1, \cdots,y^m) \sim q(\textbf{y};\textbf{x})$. 
The attacker observes $\tilde{\textbf{x}} = (\tilde{x}^1, \cdots, \tilde{x}^m)$ and chooses an exploit $e\in E$, which will target \emph{all of the devices on the defender's network}, affecting a subset of them that are vulnerable to the chosen exploit. 
The attacker and defender's optimal utility functions, after fixing the strategy of the counterpart, become
\begin{align*}
    u_a^*(q) &= \max_z \sum_{\textbf{x}} p(\textbf{x})\sum_{\textbf{y}} q(\textbf{y};\textbf{x}) \sum_e z(e;\tilde{\textbf{x}})  \sum_k v(x^k)\delta(x^k \in X^e)\\
    u_d^*(z) &= -\min_q \sum_{\textbf{x}}p(\textbf{x})\sum_{\textbf{y}} q(\textbf{y};\textbf{x})\left(\sum_e z(e;\tilde{\textbf{x}})  \sum_k v(x^k)\delta(x^k \in X^e)+c(\textbf{y})\right).
\end{align*}
The corresponding min-max problem becomes:
\begin{equation*}
\min_{q}\max_{z} \quad \sum_{\textbf{x}} p(\textbf{x})\sum_{\textbf{y}} q(\textbf{y};\textbf{x}) \left(\sum_e z(e;\tilde{\textbf{x}})  \sum_k v(x^k)\delta(x^k \in X^e) + c(\textbf{y})\right).
\end{equation*}
This can be solved using a straightforward variation of Algorithm \ref{A:gda}, with the strategy representations and loss function modified as above.

\section{Experiments}

\subsection{Near-Optimality of Generative Adversarial Masking}

Our first goal is to evaluate the quality of solutions produced by the proposed \emph{GAM} approach.
The only reliable way to do this is to compare to optimal solutions, but as we noted earlier, our lone approach for computing optimal solutions to combinatorial masking games is the LP with constraint generation (\emph{LP+CG}), which scales poorly.
Our first set of experiments, therefore, is focused on small-scale problem instances in order to evaluate how close to optimal the \emph{GAM} solutions are.
For these experiments, we let $x \in \{-1,1\}^n$ and $p(x)$ is a uniform distribution.
We let $c(y) = \sum_i 0.01 (1-y_i)$, while $v(x) = \frac{1}{2}\sum_i (x_i +1)$ (that is, the number of features that are 1).
We draw 5000 samples of $x$ and $r$ in \emph{GAM} alorithm training.
All results are averages of 100 draws of actual device configurations $\mathbf{x}$ (observable to the defender).
For each dimension $n$ in this experiment we pre-generated 2 exploits (specified in Table~\ref{cplex_gan}) for all of the runs; these constructed a prior (and not randomly generated) in order to avoid trivial solutions.
The \emph{GAM} is trained on GPU NVIDA GeForce GTX 1050 Ti using Pytorch and Cuda.
Linear programs are solved using CPLEX with tolerance set to $10^{-5}$.
Recall that $n$ denotes the number of device features while $m$ is the number of devices.

\begin{table}[h!]
\centering
\caption{Comparison between \emph{LP+CG} and \emph{GAM}.}
\label{cplex_gan}
\begin{tabular}{||c|c|c|c|c|c|c||}
\hline
\multicolumn{1}{||c}{\multirow{2}{*}{$n$}}      & \multicolumn{1}{|c}{\multirow{2}{*}{$m$}}       &
\multicolumn{1}{|c}{\multirow{2}{*}{Exploit Requirements*}}  &
\multicolumn{2}{|c|}{defender loss}             &
\multicolumn{2}{c||}{run time (seconds)}        \\ \cline{4-7} 
\multicolumn{1}{||c|}{}                         & 
\multicolumn{1}{c|}{}                           &
\multicolumn{1}{c|}{}                           & \multicolumn{1}{c|}{\emph{LP+CG}}               & \multicolumn{1}{c|}{\emph{GAM} (mean)}          & %
\multicolumn{1}{c|}{\emph{LP+CG}}               & \multicolumn{1}{c||}{\emph{GAM}}                \\ \hline
2       & 
2       & 
[-1,1],[1,-1]    &
1.52    & 
1.57 $\pm$ 0.07    & 
1.9     & 
1.7     \\
4       & 
1       & 
[-1,1,-1,-1],[1,-1,1,1]    &
1.25    & 
1.26 $\pm$ 0.03    & 
1.3     & 
1.7     \\
5       & 
1       &
[-1,1,1,-1,-1],[-1,-1,1,-1,1]    &
0.88    & 
0.90 $\pm$ 0.04    & 
5.6     & 
1.9     \\
6       & 
1       &
[-1,1,1,-1,-1,-1],[-1,-1,1,-1,1,-1]    &
1.01    & 
1.04 $\pm$ 0.06   & 
64      & 
1.9     \\ \hline
\multicolumn{7}{l}{\small * Note that we use a simpler representation here than above: 1 represents that the } \\
\multicolumn{7}{l}{\small \ \ \ configuration has to be 1, while -1 means the associated feature does not matter.}

\end{tabular}
\end{table}

The results are presented in Table~\ref{cplex_gan}, and show that the \emph{GAM} approach yields near-optimal solutions.
Moreover, even at this scale we can already observe a dramatic advantage it has in scalability: even when $n= 6$, the \emph{LP+CG} method clocks in at 64 seconds, whereas the running time of \emph{GAM} is nearly unchanged (1.7-1.9 seconds) between $n = 2$ and $n=6$.

\subsection{Systematic Large-Scale Experiments}
\label{S:exp-ls}

\subsubsection{Experiment Setup}

Our next goal is to investigate the efficacy of the \emph{GAM} approach to solving combinatorial masking games at scale, in comparison to several baselines.
Throughout, we let $c(y) = \sum_i c (1-y_i)$,
where $c$ is a constant we systematically vary in the experiments.

The feature vector $x^k$ for each device on the defender's network is constructed i.i.d. according to the following model.
The first three dimensions correspond to the Operating System installed, for which we have three options: Windows, Linux, or Mac OS.
The next three dimensions correspond to the associated versions of each of these installed, and we constrain that exactly one OS is installed, and only one version of it.
The last $50\%$ of the features correspond to ports (which may be either open, -1, or closed, 1); we constrain that at least one port is open on each device.
The remaining features correspond to applications (binary, corresponding to installed, or not) and their associated versions, with the constraint that an installed application has only a single version.
The features corresponding to versions are -1 if the associated OS/application is not installed, and integers between 1 and $V$ otherwise, where $V$ is set to either 1 (i.e., binary attributes) or 3 as specified in the experiments below.
Aside from the constraints above, each $x^k$ is generated uniformly at random (i.e., we randomly choose which OS and applications are installed, which versions of these, and which ports are open).
We set $v(x)=1+$\emph{[the number of installed applications]}.

We construct the set of exploits (the size of which we systematically vary) as follows.
Each exploit $e \in E$ either targets a contiguous sequence of versions of a particular OS, or a contiguous sequence of versions of both an OS and an application.
In either case, a particular target port is required to be open for the exploit to succeed.
All choices above are made uniformly at random.
For each experimental setting, we tune the parameters of \emph{GAM} through a pre-testing phase.
In all cases, \emph{GAM} takes $10^4$ samples of $x$ and $r$, and the results provided are averages of $10^3$ runs.
In the experiments we use GPU NVIDIA TITAN Xp, GeForce RTX 2080Ti, and GeForce GTX 1080Ti using Pytorch and Cuda.

\subsubsection{Baseline Approaches}

We compare \emph{GAM} to the following three baselines:

\smallskip
\noindent{\it Random Masking } This is a simple baseline in which the mask is chosen uniformly at random from $\{0,1\}^n$ for any $\mathbf{x}$.
Since the run time of this is negligible, we do not report it below.

\smallskip
\noindent{\it Unconditional Masking } A natural baseline is to use a simpler mixed strategy for the defender $q(\mathbf{y};\mathbf{x})$ which is independent of $\mathbf{x}$, i.e., $q_{\mathit{unc}}(\mathbf{y})$.
We can still apply a simplified version of the \emph{GAM} approach to compute the associated distribution.

\smallskip
\noindent{\it Greedy Masking } Greedy search generates the mask $y$ by iteratively minimizing the expected marginal loss.
We initialize $y$ as not masking any configurations.
In each step, we decide whether to mask an (additional) attribute that has not yet been masked, choosing an attribute that yields than greatest reduction in expected loss to the defender (and stopping if this is negative).
Note that here we also assume that the masking strategy $y$ is independent of $x$. 
Even in this case, greedy masking is time consuming, since in order to evaluate the marginal impact of masking we need to execute the attacker's best response, which itself entails training the best response neural network $z$.

\subsubsection{Results}

We begin by comparing \emph{GAM} with the \emph{unconditional masking} baseline in which $q_{\mathit{unc}}(\mathbf{y})$ does not depend on the true state $\mathbf{x}$ (in game-theoretic language, this would correspond to a \emph{pooling strategy}, which is uninformative as regards to the true device attributes).
In this experiment, we consider a single device, fix $c = 0.05$ and set $V=3$, and further simplify by considering a single possible version for each OS and application.

\begin{figure}[h!]
\centering
\includegraphics[width=0.5\textwidth]{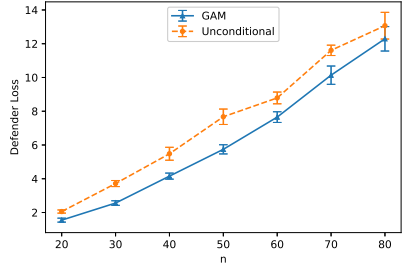}
\caption{Comparison of defender loss between \emph{GAM} and \emph{Unconditional Masking}.}
\label{fig:loss_errorbar}
\end{figure}
The results are shown in Figure \ref{fig:loss_errorbar}.
We can observe that \emph{GAM} is significantly better than unconditional masking over a range of $n$, with the difference often above 25\%.
This demonstrates that a simple \emph{pooling} strategy is inadequate, and it is critical to condition the mixed strategy of the defender on the true state of the devices.

Next, we run four sets of experiments and compare to our \emph{GAM} approach to the remaining two baselines (random and greedy masking) in terms of both defender loss and running time.

In the first experiment, we systematically vary $n$, with the number of exploits matching the dimension.
We keep $m=1$, and set cost $c=0.01$.
We set $V=3$.
\begin{figure}[h!]
\centering
\begin{subfigure}{.48\textwidth}
  \centering
  \includegraphics[width=1.0\linewidth]{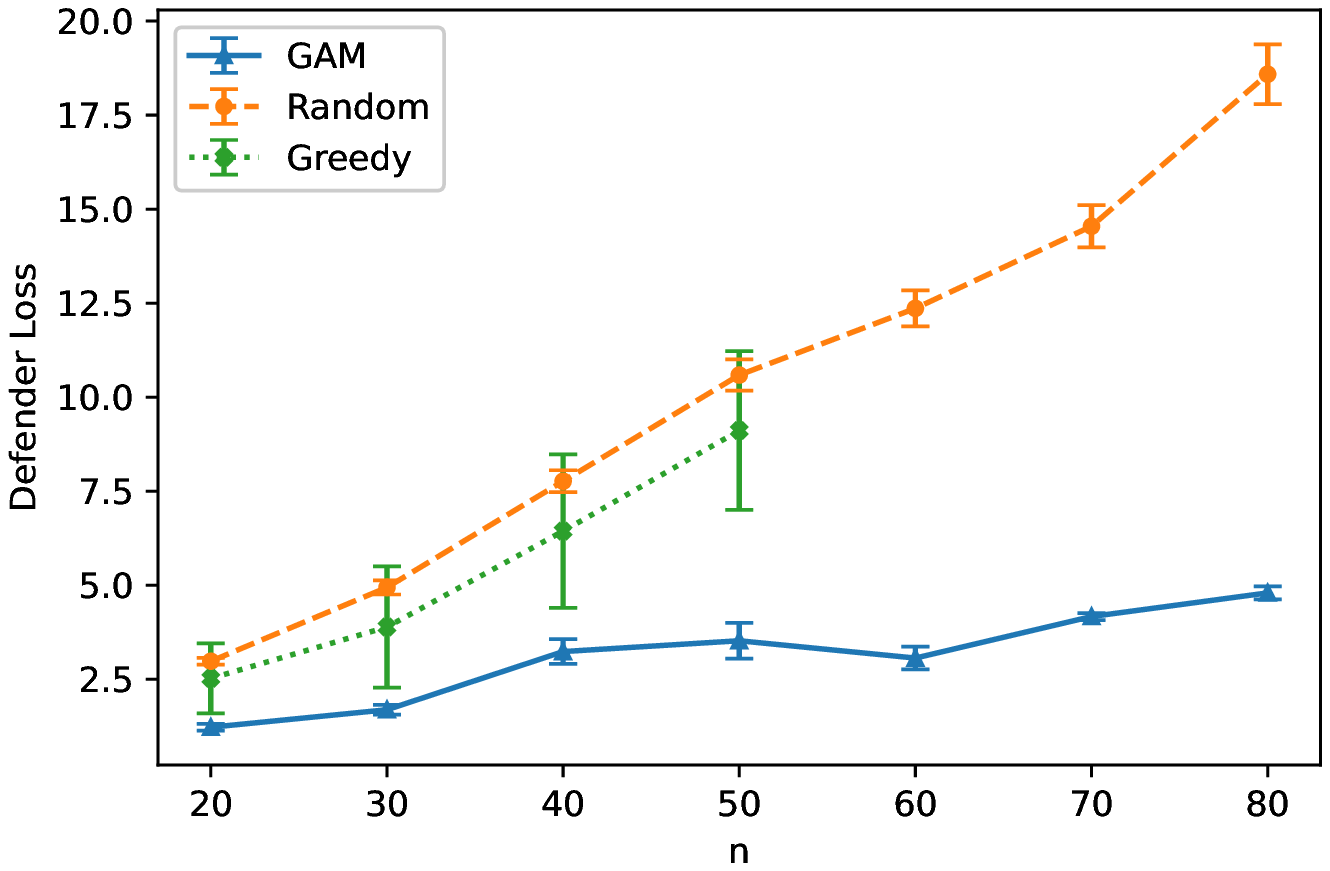}
\end{subfigure}%
\begin{subfigure}{.48\textwidth}
  \centering
  \includegraphics[width=1.0\linewidth]{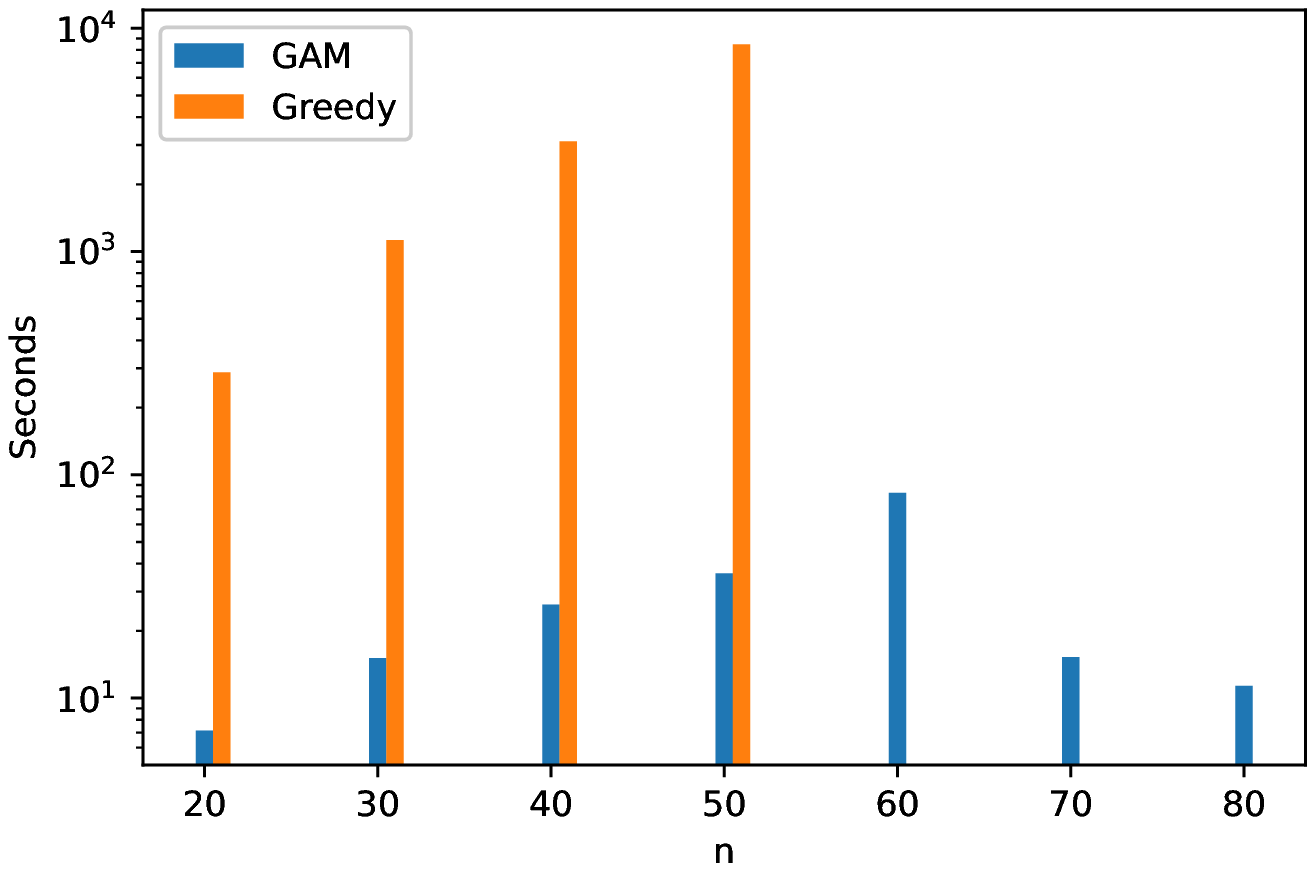}
\end{subfigure}
\caption{Experiment 1 results: efficacy and scalability as a function of $n$.}
\label{fig:e1}
\end{figure}
Figure \ref{fig:e1} (left) shows that \emph{GAM} significantly outperforms the two baselines in terms of defender loss, particularly as we increase the number of attributes $n$.
Interestingly, Greedy is only slightly better than Random, and it fails to scale beyond $n=50$.
Figure \ref{fig:e1} (right) compares the running time between \emph{GAM} and Greedy, demonstrating that while \emph{GAM} is remarkably scalable, with little difference in running time between $n=20$ and $n=80$, Greedy is significantly slower, and fails to scale well with $n$.

In Experiment 2, we fix $n=20$, $m=1$, $c=0.01$, and $V=3$, and systematically vary the number of exploits.
\begin{figure}[h!]
\centering
\begin{subfigure}{.48\textwidth}
  \centering
  \includegraphics[width=1.0\linewidth]{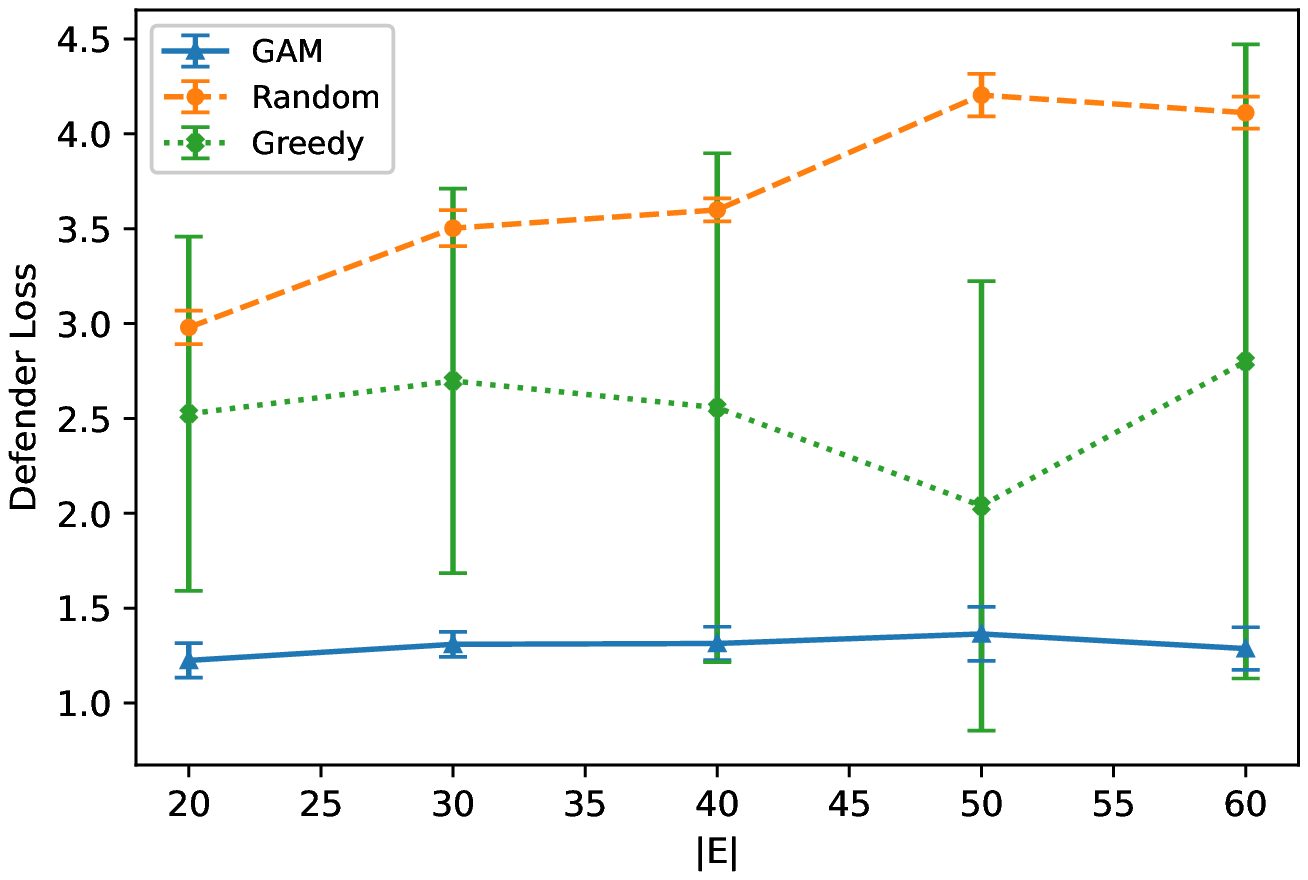}
\end{subfigure}%
\begin{subfigure}{.48\textwidth}
  \centering
  \includegraphics[width=1.0\linewidth]{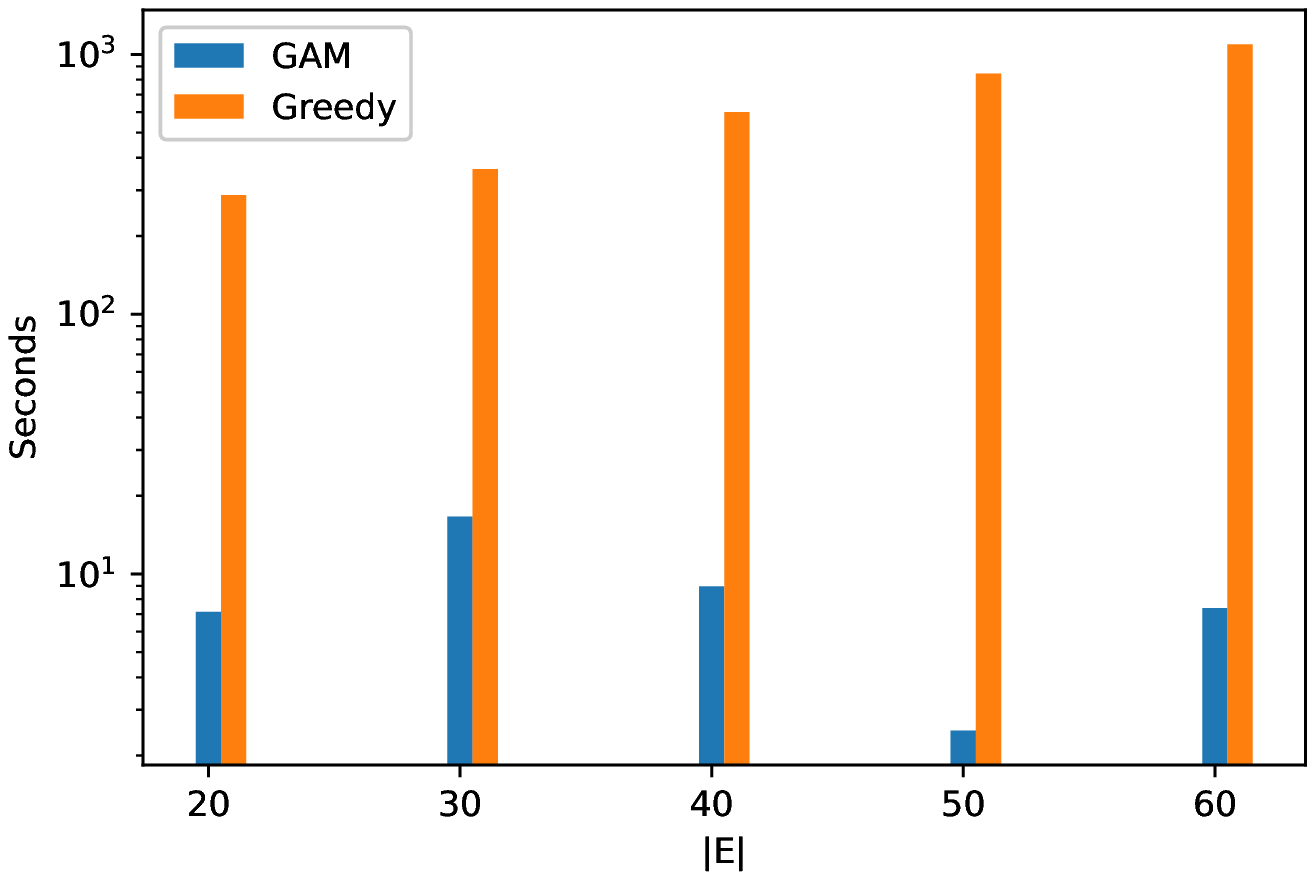}
\end{subfigure}
\caption{Experiment 2 results: efficacy and scalability as a function of $|E|$.}
\label{fig:e2}
\end{figure}
Figure~\ref{fig:e2} shows again that \emph{GAM} significantly outperforms both baselines in terms of defender's loss, and Greedy in terms of running time.
It is interesting to note that the defender's loss in all three approaches appears to depend only weakly on the number of exploits available.
This demonstrates the value of deception: although increasing the number of exploits also increases the likelihood that at least one can successfully attack the defender's device, deception serves to make it difficult for the attacker to choose the correct one.

Experiment 3 now systematically varies the relative cost $c$ of masking, keeping $n=20$, $m=1$, $V=1$, and $|E|=20$.
\begin{figure}[h!]
\centering
\begin{subfigure}{.48\textwidth}
  \centering
  \includegraphics[width=1.0\linewidth]{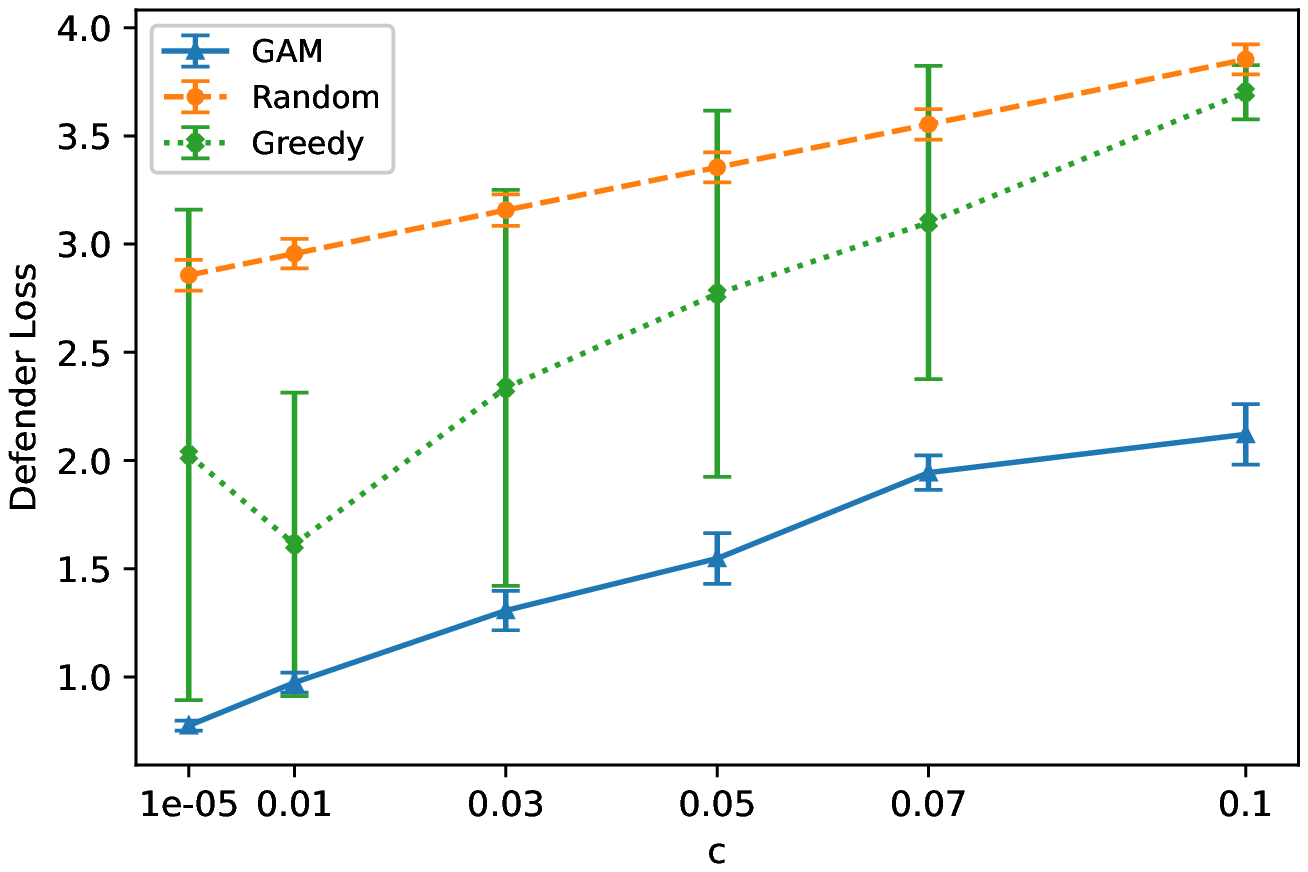}
\end{subfigure}%
\begin{subfigure}{.48\textwidth}
  \centering
  \includegraphics[width=1.0\linewidth]{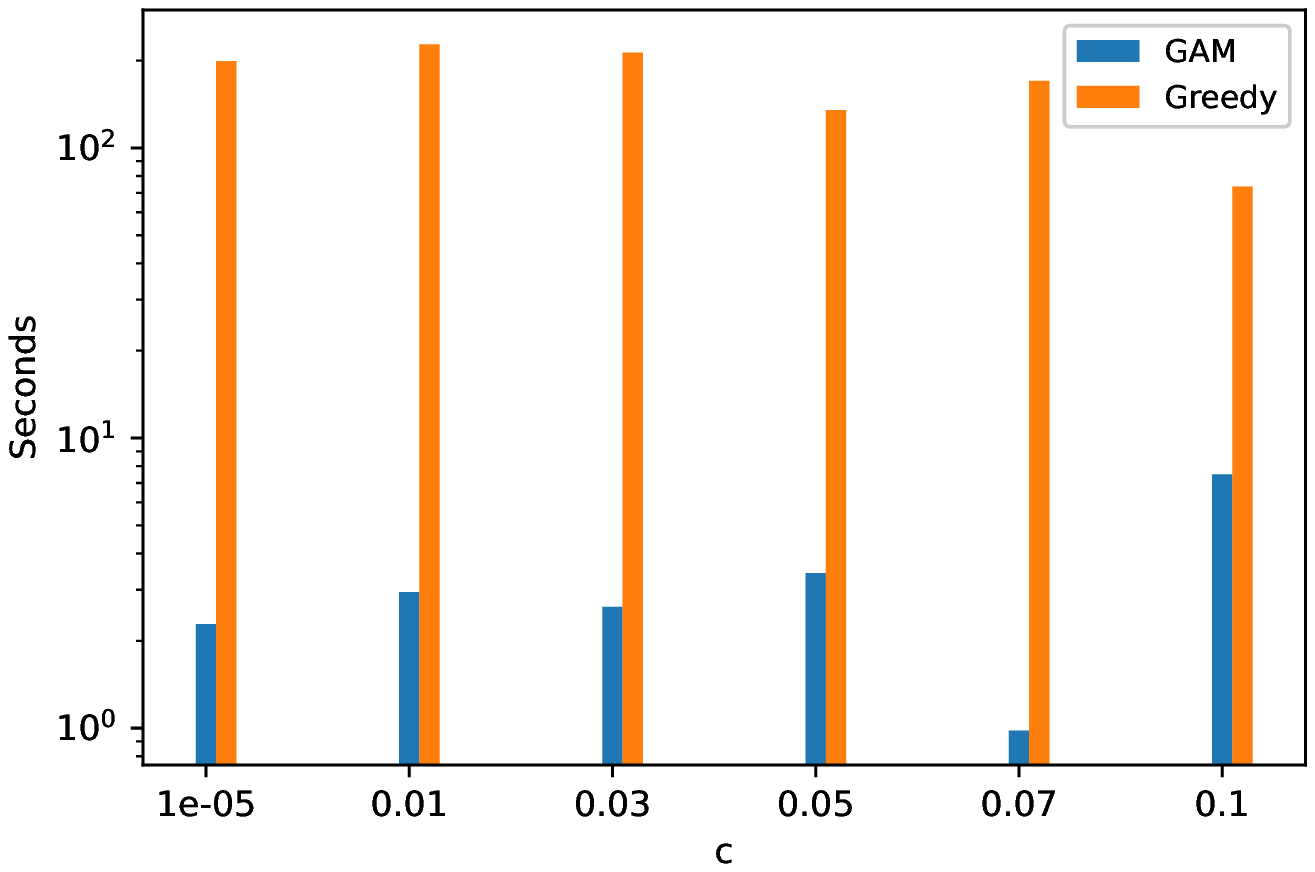}
\end{subfigure}
\caption{Experiment 3 results: efficacy and scalability as a function of $c$.}
\label{fig:e3}
\end{figure}
Figure~\ref{fig:e3} presents the results.
As we can expect, defender's loss increases as we increase $c$ (as masking becomes more expensive), but \emph{GAM} remains significantly better than the baselines.
Running time (Figure~\ref{fig:e3}, right) again shows a significant advantage of \emph{GAM} over Greedy.

\begin{figure}[h!]
\centering
\begin{subfigure}{.48\textwidth}
  \centering
  \includegraphics[width=1.0\linewidth]{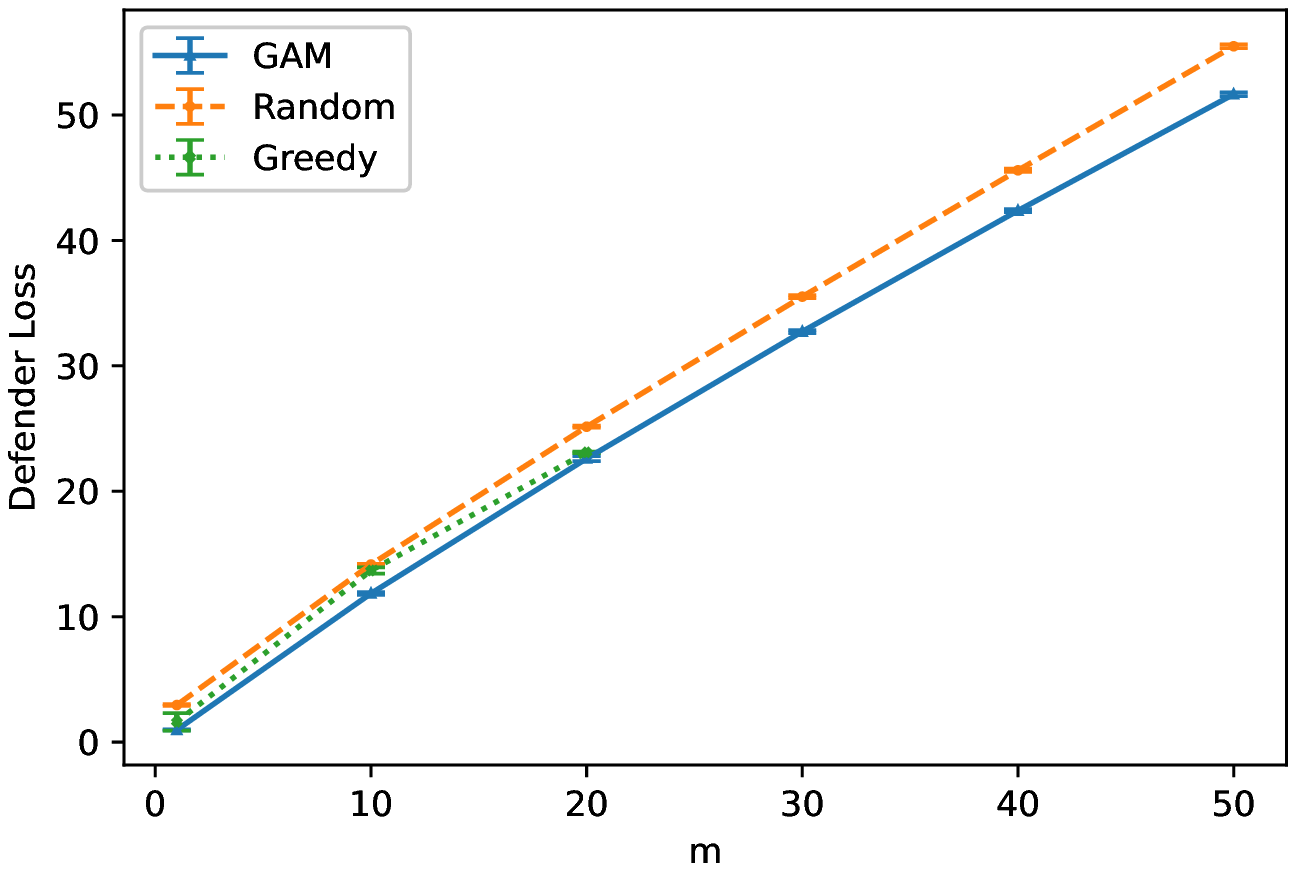}
\end{subfigure}%
\begin{subfigure}{.48\textwidth}
  \centering
  \includegraphics[width=1.0\linewidth]{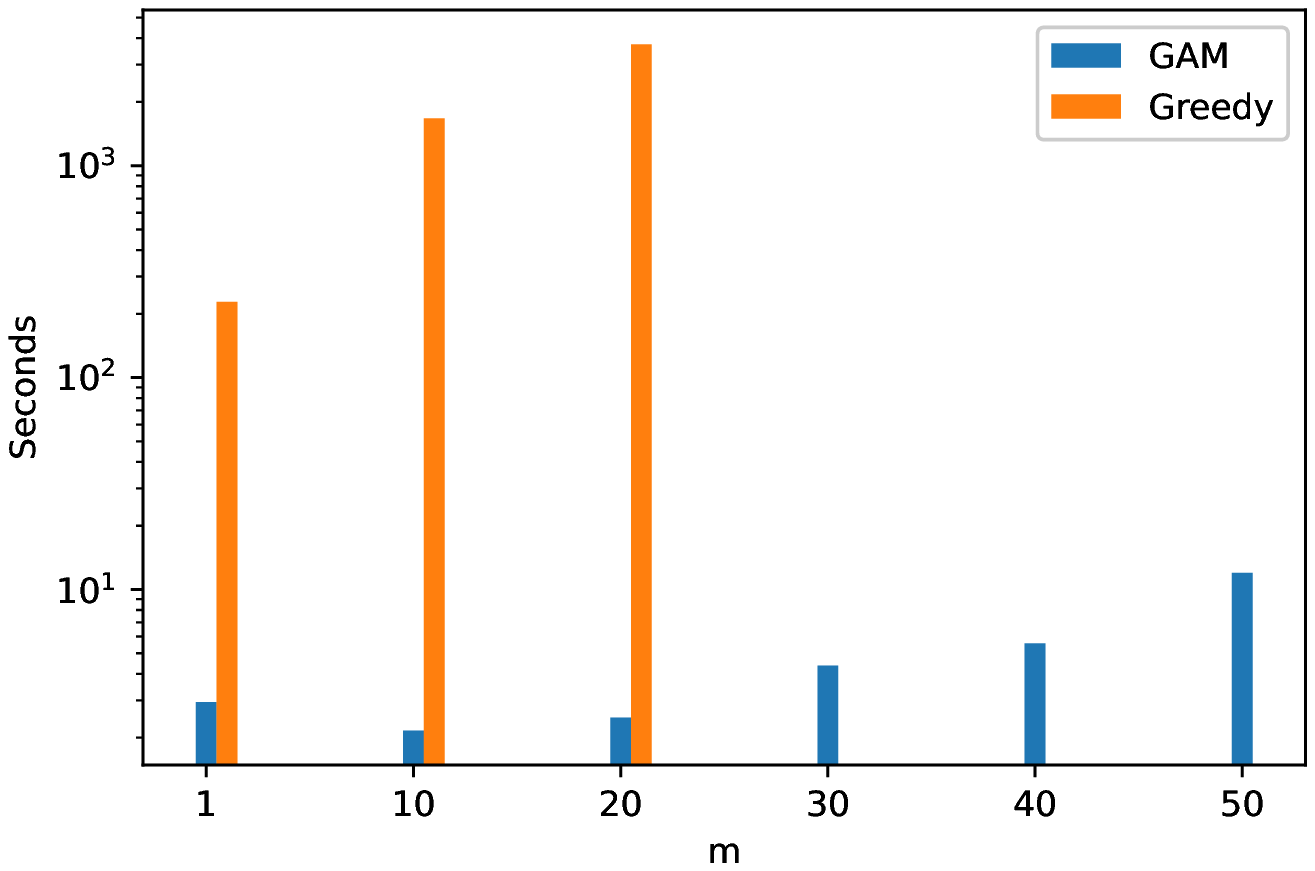}
\end{subfigure}
\caption{Experiment 4 results: efficacy and scalability as a function of $m$.}
\label{fig:e4}
\end{figure}
Finally, in Experiment 4 we study the impact of the number of devices $m$.
We set $n=20$, $c=0.01$, $V=1$, and $|E|=20$ for this experiment.
As shown in Figure~\ref{fig:e4} (left), the defender's loss scales roughly linearly (as one would expect) with the number of devices, with \emph{GAM} still offering the best solution.
Scalability results (Figure~\ref{fig:e4}, right) show again that \emph{GAM} scales much better than Greedy. While Greedy could not run for high values of $m$, the number of devices $m$ appears to have relatively limited impact on the running time for \emph{GAM}, which grows from $\sim 3$ seconds to $\sim 12$ seconds as $m$ increases from 1 to 50.

\section{Case Study}

We now use a case study to delve more deeply into the nature of masking and attack strategies obtains as solutions to the \emph{CGM}.
We consider a single device, and use the same $v(x)$ and $c(y)$ as above, with $c=0.01$.
We let $n=20$, generating $x$ as described in Section~\ref{S:exp-ls}, and generate 19 exploits with requirements documented in Table~\ref{exp}.
\begin{table}[h!]
\centering
\caption{Exploit requirements.}
\label{exp}
\begin{tabular}{||r|r|r|r|r|r||}
\hline
Exploit \# & System & System Version & App & App Version & Port \\
\hline
1  & 0      & 2, 3           & 3   & 1           & 19   \\
2  & 0      & 2              & 8   & 1, 2, 3     & 14   \\
3  & 0      & 2, 3           &     &             & 17   \\
4  & 1      & 2, 3           & 3   & 1, 2        & 14   \\
5  & 1      & 1              & 4   & 1, 2        & 12   \\
6  & 1      & 1, 2, 3        & 4   & 3           & 17   \\
7  & 1      & 1, 2, 3        & 5   & 1, 2, 3     & 14   \\
8  & 1      & 3              & 5   & 3           & 19   \\
9  & 1      & 2              & 6   & 1           & 12   \\
10 & 1      & 1, 2           & 7   & 2           & 11   \\
11 & 1      & 2, 3           & 7   & 1, 2, 3     & 17   \\
12 & 1      & 3              &     &             & 16   \\
13 & 2      & 2, 3           & 3   & 1, 2        & 12   \\
14 & 2      & 2, 3           & 5   & 1, 2        & 11   \\
15 & 2      & 1, 2           & 5   & 3           & 12   \\
16 & 2      & 1, 2           & 6   & 1, 2, 3     & 12   \\
17 & 2      & 1, 2, 3        & 8   & 1, 2, 3     & 18   \\
18 & 2      & 1, 2, 3        & 9   & 1, 2, 3     & 13   \\
19 & 2      & 1, 2, 3        &     &             & 10  \\
\hline
\end{tabular}
\end{table}

\begin{figure}[h!]
\centering
  \centering
  \includegraphics[angle=90, width=0.75\textwidth]{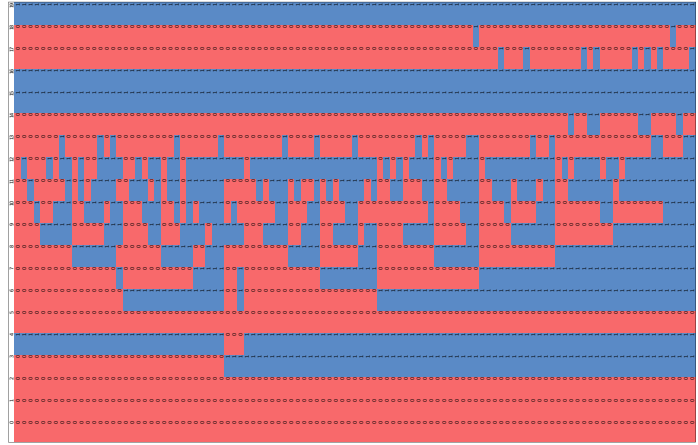}
\caption{Visualization of the support of the masking strategy computed by \emph{GAM} in the case study. Red encodes a decision to mask an attribute, while blue encodes a decision not to mask.}
\label{fig:cs-strat}
\end{figure}

Figure \ref{fig:cs-strat} visualizes the BNE masking strategy, with the support (masks chosen with positive probability) as columns, and device attributes as rows.
Red corresponds to attributes chosen to be masked, while blue encodes a decision not to mask an attribute.
We can glean several insights from this figure.
First, note that we always mask OS information. This is because all exploits have specific OS requirements.
Moreover, exploits 3, 12, and 19 \emph{only} have an OS requirement (i.e., they exploit the OS vulnerabilities, rather then application vulnerabilities).
The masking strategy also always masks application 5, because there are 4 exploits available targeting this application (exploits 7, 8, 14, 15), the most of any applications, and these become obvious choices if application 5 is known to be installed.
Note, moreover, we never mask whether port 15 is open, as no exploits target it.

\begin{table}[h!]
\centering
\caption{Attacker's strategy.}
\begin{tabular}{||r|r|r|r|r|r|r||}
\hline
Exploit \# & Avg Pctg & System & System Version & App & App Version & Port \\
\hline
19         & 85.7\%       & 2      & 1, 2, 3        &     &             & 10   \\
7          & 5.3\%        & 1      & 1, 2, 3        & 5   & 1, 2, 3     & 14   \\
18         & 5.3\%        & 2      & 1, 2, 3        & 9   & 1, 2, 3     & 13   \\
3          & 2.2\%        & 0      & 2, 3           &     &             & 17   \\
17         & 1.5\%        & 2      & 1, 2, 3        & 8   & 1, 2, 3     & 18  \\
\hline
Sum & 100.0\%& & & & &\\
\hline
\end{tabular}
\label{tab:attacker}
\end{table}
Finally, we study the attacker's mixed strategy by computing the \emph{average} probability of choosing each exploit over all possible attacker observations.
Table \ref{tab:attacker} shows the results for exploits with a non-negligible probability of being chosen.
Note that one exploit (19) is chosen an overwhelming fraction of the time, and only 5 are chosen with any frequency.
All 5 exploits chosen have relatively weak requirements: 19 and 3 only require a particular OS and port (these exploits have the weakest requirements of all), while 7, 17, and 18 can exploit any extant version of their targeted application.

\section{Conclusion}

This paper studies a form of cyber deception in which a defender can only mask a subset of observable device attributes, while the attacker chooses one of a collection of exploits to deploy against the defender's network.
The strategic interaction can be viewed as a combinatorial signaling game in which the defender's privacy information (a vector of attributes, or features, about their devices) is partially and strategically leaked to the adversary, with the express goal of deception in order to make it difficult for the adversary to carefully target exploits against the network.
The adversary, in turn, reasons about such deception to choose an exploit that maximizes their expected posterior utility.
Solving this game exactly for a Bayes-Nash equilibrium becomes rapidly intractable for even a small number of attributes.
We address this challenge by first encoding the strategies of both players as neural networks, with the defender's mixed strategy represented by a conditional generative neural network.
We then propose a gradient-based approach for learning the approximate equilibrium solutions of the game.
Our experiments show that the proposed approach is highly effective and highly scalable, while a case study confirms that it yields intuitive solutions to realistic cybersecurity encounters.

\section*{Acknowledgments}

This work was partially supported by the National Science Foundation (IIS-1905558 and ECCS-2020289) and Army Research Office (W911NF1910241 and W911NF1810208).

\bibliographystyle{splncs04}
\bibliography{main}

\end{document}